\documentclass[epj,twocolumn]{webofc}
\usepackage[varg]{txfonts}   % Web of Conferences font
\usepackage{color}
\usepackage{graphicx} 
\usepackage{amssymb,amsmath,enumerate} 
\begin{document}
\title{12 GeV CEBAF - The Physics and Experiments}
\author{\firstname{Volker~D.} \lastname{Burkert}\inst{1}\fnsep\thanks{\email{burkert@jlab.org}}, for the CLAS collaboration }
\institute{Thomas Jefferson National Accelerator Facility \\ 12000 Jefferson Avenue, Newport News, Virginia, USA  } 
\abstract{  In this talk the role of spin and polarization is discussed in experimental search for new 
excited baryon states and in the study of the internal quark-gluon structure of the proton and neutron.  
 Also the perspective of spin physics at the 12 GeV CEBAF electron accelerator is discussed and what 
 we hope to learn about fundamental properties of hadrons such as their multi-dimensional structure through 
 the momentum and spatial imaging, and about the forces on the quarks in the proton and how 
 quark confinement may be realized through the spatial distribution of such forces.}
\maketitle
\section{Introduction}
\label{intro}
\begin{figure}[h]
\includegraphics[width=8.0cm,height=6.5cm,clip]{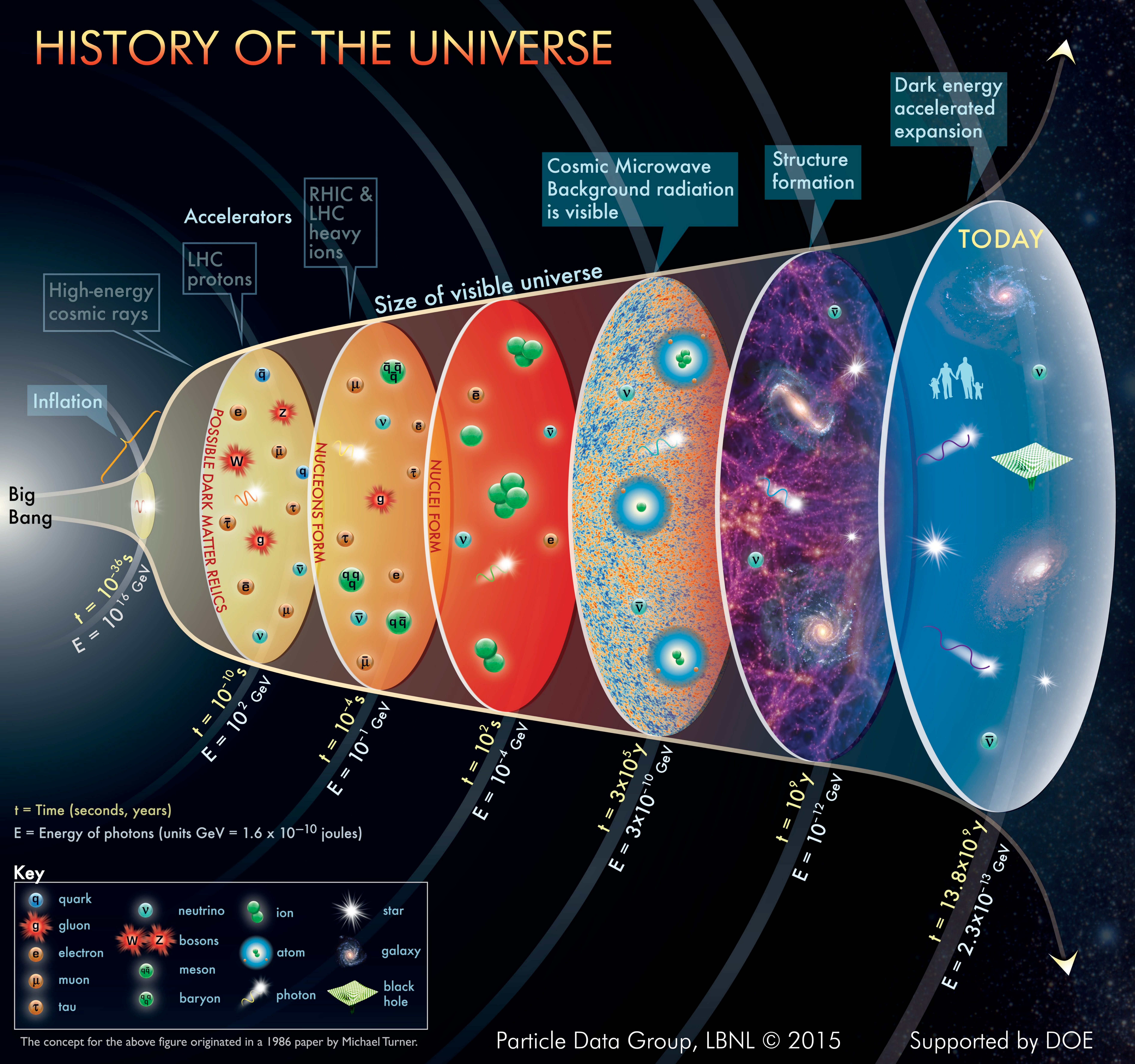}
\caption{The evolution of the Universe as pictured by the LBNL Particle Data Group, 2015. 
The area characterized by the second disk from the left is where hadron of confined quarks and gluons occur. The CEBAF electron accelerator has the energy reach to access this region and study processes in isolation that occurred in the microsecond old universe and resulted in the freeze out of baryons isolation. }
\label{universe}      
\end{figure}

Dramatic events occurred in the evolution of the microsecond old universe that had tremendous implication for 
the further development of the universe to the state it is in today. As the universe expanded and cooled sufficiently
into the GeV range (see Fig.~\ref{universe}), the transition occurred from the phase of free (unconfined) quarks and 
gluons, to the hadron phase with quarks and gluon confined in volumes of $\approx 1$fm$^3$, i.e. protons, neutrons, and 
anti-baryons. In course of this process, elementary, nearly massless quarks acquire dynamical mass due to 
the coupling to the gluon field, and chiral symmetry is broken. 
This transition is not a simple first order phase transition, but a "cross over"  between two phases 
of strongly interacting matter, which is moderated by the excitation of baryon resonances starting 
from the highest mass states and ending with the low mass states. 

A quantitative understanding of this transition requires more excited baryons of all flavors, than have been observed 
to date and are included in the "Review of Particle Properties"~\cite{Bazavov:2014xya,Bazavov:2014yba}. These 3 
phenomena, the presence of the full complement 
of excited baryons, the acquisition of dynamical mass by light quarks, and the transition from unconfined quarks to 
confinement are intricately related and are at the core of the problems we are trying to to solve in hadron physics today. 
Although we cannot recreate the exact conditions that existed during this phase in the history of the universe, we 
do have all the tools at our disposal to search for new states and study the individual states in relative isolation, by 
probing the quark mass versus the momentum or distance scale, and study the forces in the nucleon, which provide the 
confinement of quarks and gluons in the nucleon.

Accounting for excitation spectrum of protons and nucleons and describing the effective degrees of freedom is one of the 
most important and certainly the most challenging task in hadron physics. It has been the focus of the CLAS N* 
program at Jefferson Lab and likely will remain so with its extensions towards higher energies with CLAS12. 
The first part of the talk is focussed on what has been achieved in discovering new excited baryons, and in revealing the active degrees of freedom. In the second part I discuss experiments that are planned to explore the internal quark structure and the quark longitudinal and transverse spin distributions as well as what we could learn about the confinement forces underlying 
the internal quark structure. 

\section{The CW Electron Accelerator and the 12 GeV experimental equipment. } 
\label{equip}
\begin{figure}[ht]
\includegraphics[width=8cm,height=8cm]{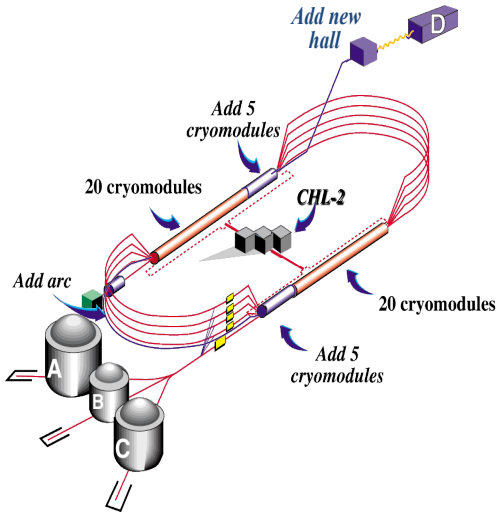}
\vspace{-0.7cm}
\caption{The Jefferson Lab continuous electron beam accelerator facility showing the components needed for the 12 GeV upgrade.} 
\label{cebaf}    
\end{figure}

\noindent The electron accelerator is shown schematically in Fig.~\ref{cebaf}. The two linear accelerators are based on superconducting rf technology. Electrons are injected into an injector shown at the upper left end of the racetrack, where they are pre-accelerated, injected into the north linac and boosted in energy to  600 MeV. A set of magnetic dipole magnets deflects the beam first out of plane into the uppermost of the recirculating arcs where the beam is bent by 180 degrees and injected into the south linac to be accelerated to up to 1200 MeV. This is repeated four more times to reach the final energy of 6000 MeV. For the energy upgrade five accelerating cryomodules with four times higher gradients per unit length are added to both linacs to reach a maximum energy at the existing end stations of nearly 11 GeV. One arc and one more path through the north linac are added to accelerate the beam to 12 GeV and transport it to the new Hall D. 

Hall D, which houses the GlueX experiment, a 4$\pi$ solenoid based detector uses a linearly polarized photon beam, initially 
mostly for meson spectroscopy experiments. Hall C has a pair of focussing high momentum spectrometers, an existing HMS and 
the new one under construction super-high momentum spectrometer (SHMS) for use in precision coincidence measurements, for 
L/T separations among other applications. Hall A will add a Super Bite Spectrometer (SBS) to the existing High Resolution 
Spectrometers. The focus is on elastic form factor measurements, and SIDIS experiments with polarized He-3 target. Hall B 
is in the process of completing the installation of CLAS12, a new and upgraded version of CLAS with a large superconducting 
Torus magnet and forward tracking and particle identification, and a solenoid magnet for central charged particle tracking and id.  
A new polarized solid state target with NH3 and ND3 as target material is under development fro use in CLAS12.   

In the next section I will discuss results on baryon spectroscopy from the 6 GeV operation of CEBAF where polarization
was essential for major advances in this sector of hadron physics towards a better understanding of effective degrees of
freedom underlying the nucleon excitation spectrum.  

\section{Establishing the light quark baryon spectrum}
\label{baryon_spectrum}
Obtaining an accurate account of the full nucleon resonance spectrum is the basis for making progress in our understanding 
of strong QCD as it relates to light quark sector. To accomplish this, we need precise measurements of the nucleon excitation spectrum 
to test our best models. The symmetric quark model provides a description of the lower mass spectrum in terms of isospin
and spin-parity quantum numbers, but masses are off, and there are many states predicted within the $SU(6)\otimes O(3)$
symmetry that are missing from the observed spectrum.  Although we have already the correct theory, QCD, we cannot really test it on the nucleon spectrum, because the full spectrum is not known, and the theory is currently not in a position to predict 
more than what the quark model already has done. Our task is therefore two-fold: 1) to establish the experimental 
nucleon spectrum, and 2) to develop applications of strong QCD to be able to reliably explain the spectrum in detail, including masses and hadronic and electromagnetic couplings.       

\begin{figure}[h]
\vspace{-0.5cm}
\hspace{-0.5cm}
\includegraphics[width=9.0cm,height=8.0cm,clip]{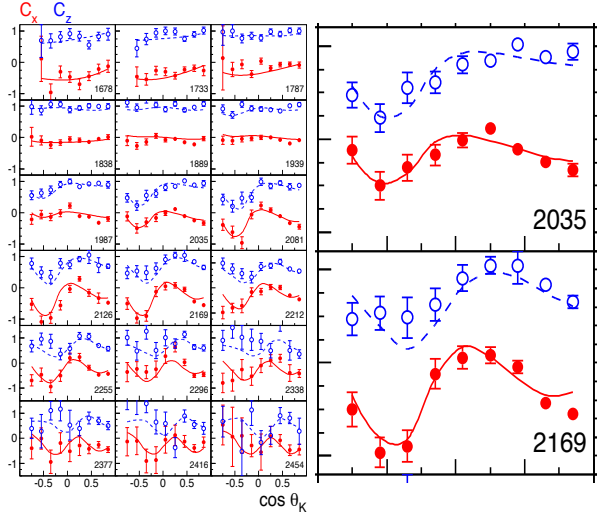}
\vspace{-1.0cm}
\caption{Left panel: Polarization transfer coefficients from the spin polarized photon to the recoil $\vec\Lambda$. 
Right panel: Two magnified bins. Solid points are the transverse component C$_x$, open circles: longitudinal 
component C$_z$. The curves 
represent fits by the Bonn-Gatchina group~\cite{Anisovich:2011fc} to the CLAS data. Other hyperon cross 
section and spin asymmetry measurements~\cite{McCracken:2009ra} were fitted simultaneously and resulted 
in the discovery of 2 new excited 
nucleon states and led to significantly higher evidence for 6 other states. }
\label{KLam_CxCz}
\end{figure}

The experimental part has been the goal of the $N*$ program with CLAS detector and with other facilities, 
especially CB-ELSA. Significant progress has been made in recent years that also  included development of 
multi-channel partial wave analysis frameworks. Much of recent progress came as a result of precise data, including measurement of polarization observables collected in the strangeness channel. Figure~\ref{KLam_CxCz} shows
examples of double polarization measurement in $\Lambda$ hyperon photoproduction~\cite{Bradford:2006ba} 
with the CLAS detector. The Bonn-Gatchina group has claimed a set of eight states that are either newly discovered or have significantly improved 
evidence for their existence.  These states entered in recent editions of the Review of Particle Properties~\cite{Olive:2016xmw}.

\subsection{Hyperon photoproduction}
\label{hyperon}
 
Here one focus has recently been on precision measurements of the $\gamma p \to K^+\Lambda$ and 
$\gamma p \to K^+\Sigma^\circ$ differential cross section~\cite{Bradford:2006ba,McCracken:2009ra}, 
and using polarized photon beams, with circular or linear 
polarization~\cite{McNabb:2003nf,Bradford:2005pt,McCracken:2009ra,Bradford:2006ba,Dey:2010hh}, 
several polarization observables can be 
 measured by analyzing the weak decay of the recoil $\Lambda \to p \pi^-$, and $\Sigma^\circ \to \gamma \Lambda$. 
 Samples of the data are shown in Fig.~\ref{KLam_CxCz}. 
 It is well known  that the energy-dependence of 
 a partial-wave amplitude for one particular channel is influenced by other reaction 
channels due to unitarity constraints. To fully describe the energy-dependence 
of an amplitude one has to include other reaction channels in a coupled-channel approach. 
Such analyses have been developed by the Bonn-Gatchina group~\cite{Anisovich:2011fc}, 
at EBAC~\cite{JuliaDiaz:2007fa}, by the Argonne-Osaka group~\cite{Kamano:2013iva}, 
and the Bonn-J\"ulich group~\cite{Ronchen:2014cna}.   

\begin{figure}[t]
\hspace{-0.2cm}\includegraphics[width=8.cm,height=7.5cm,clip]{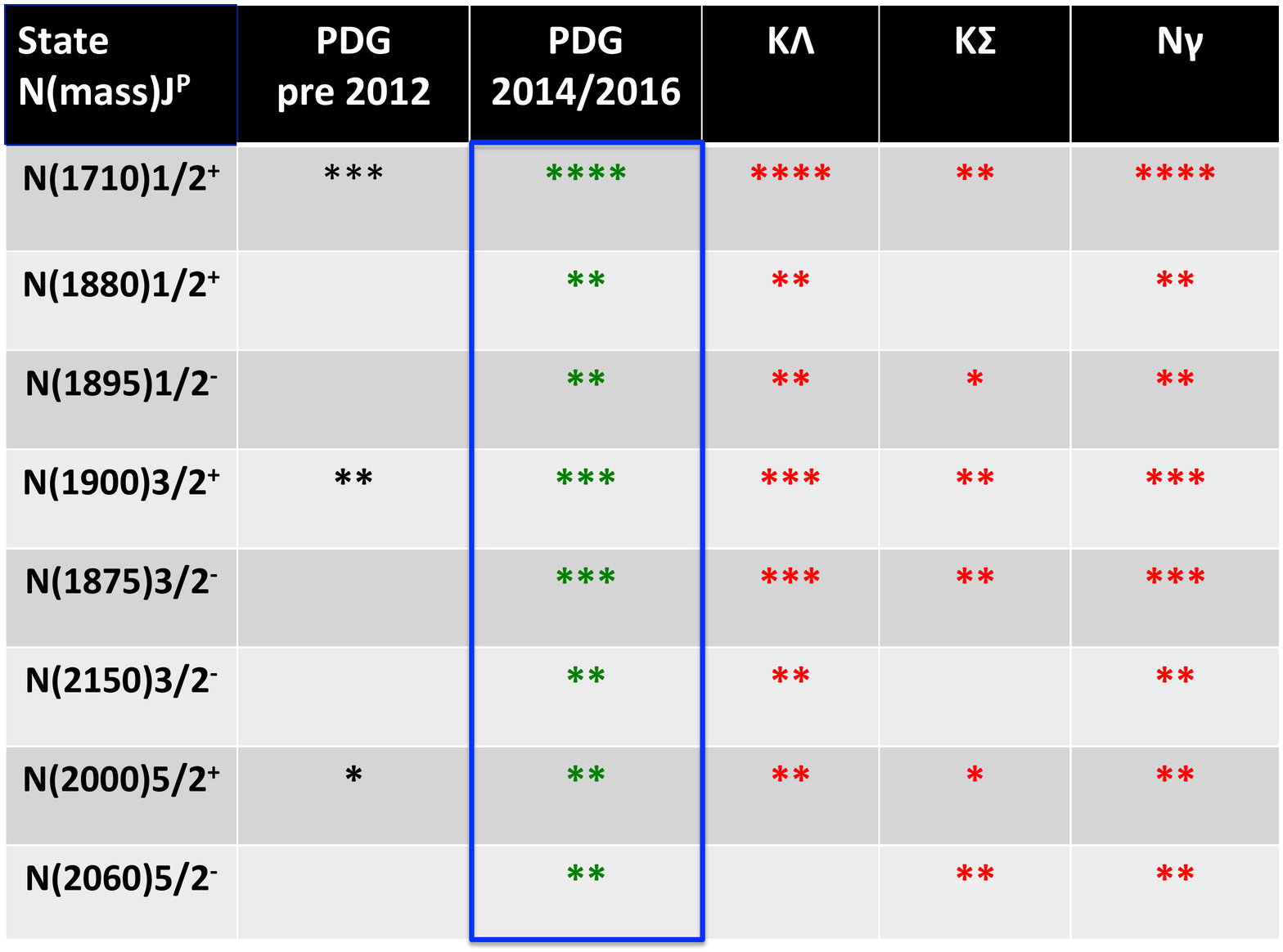}
\vspace{-1cm}\caption{Recently discovered nucleon resonances with their star ratings in the Review of Particle Properties, 2016. 
The states have been observed in the Bonn-Gatchina multi-channel partial wave analysis of photo-produced 
$KY$ final states~\cite{McCracken:2009ra,Dey:2010hh}. Columns 4 to 6 shows the relative importance of the 
$K^+\Lambda$, $K^+\Sigma^\circ$ and $\gamma p$ couplings in the observation of these states. }
\label{pdg2016}      
%\end{figure}
%\begin{figure}[hb]
\hspace{-0.3cm}\includegraphics[width=9.0cm,height=8.cm,clip]{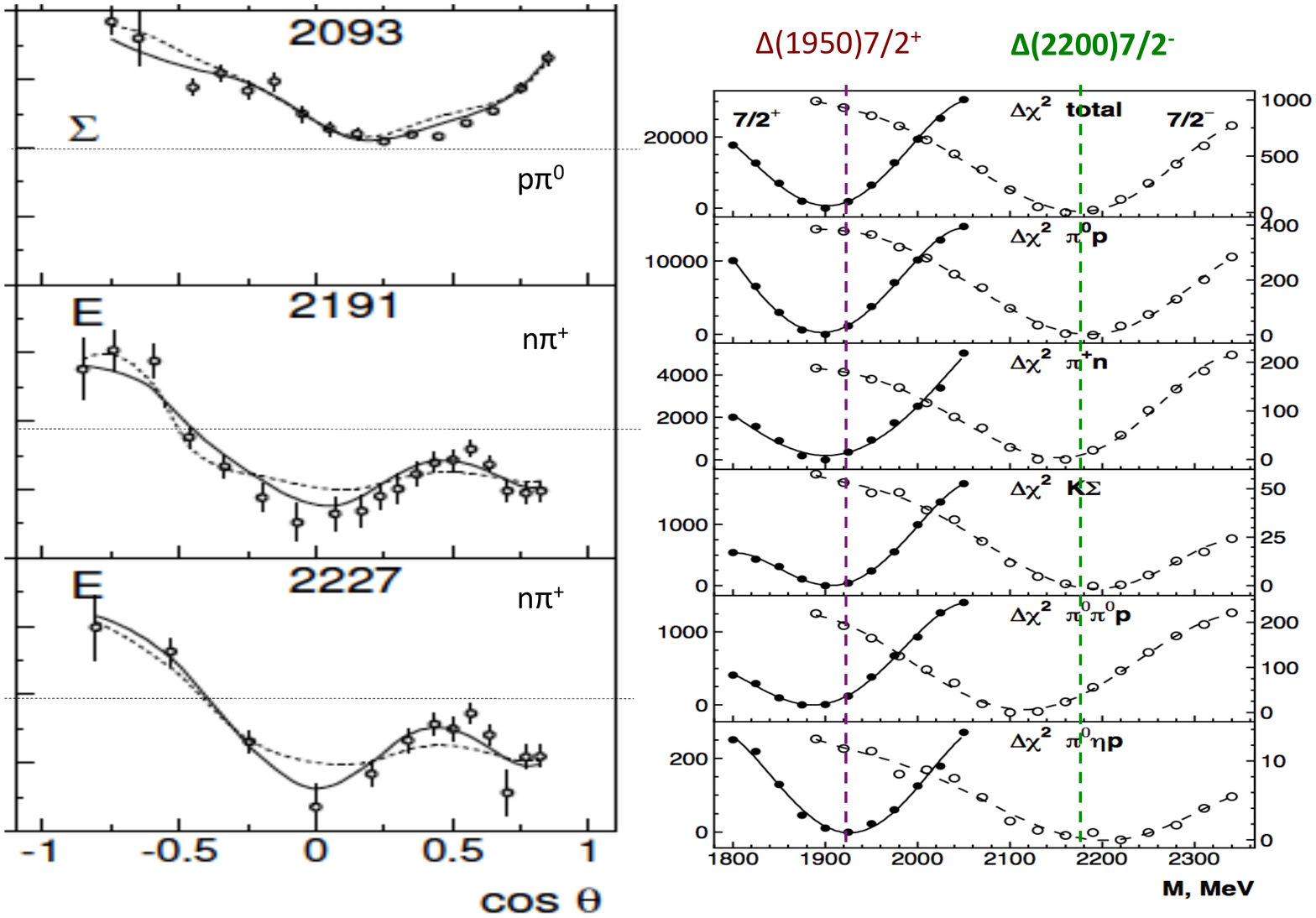}
\caption{Left panel: Beam polarization asymmetry $\Sigma$ and double spin asymmetry $E$ observables measured 
in photoproduction of $p\pi^\circ$ and $n\pi^+$, respectively.  The two curves show the Bonn-Gatchina fit 
without (dashed) and with (solid) the $\Delta(2200)$ state. Right panel: The mass scan of the data included in the fit show
the strong well-known $\Delta(1950){7\over 2}^+$ (left) and the poorly established 1* state $\Delta(2200){7\over 2}^-$ (right). The dashed lines show the approximate mass values of the Breit-Wigner fits. }
\label{Delta2200}
\end{figure}

The Bonn-Gatchina group has claimed a set of eight states that are either newly discovered or have significantly improved 
evidence for their existence.  These states entered in recent editions of the Review of Particle 
Properties~\cite{Olive:2016xmw}. 
Figure~\ref{pdg2016} shows the nucleon resonances observed in the Bonn-Gatchina multi-channel partial wave analysis
using the hyperon photoproduction data from CLAS and other data sets. 

New data on $K^+\Lambda$ production using a linearly polarized photon beam and measuring the $\Lambda$ 
recoil polarization along the $\Lambda$ momentum~\cite{Paterson:2016vmc} have been published. 
They data show strong sensitivity to excited baryon states, including possible new states, but they have 
not been included in previous full multi-channel partial wave analyses.  

%\begin{figure}[thb]
%\hspace{-0.5cm}\includegraphics[width=9.0cm,height=8.5cm,clip]{KLambda_Oz.pdf}
%\caption{Double polarization observable $O_z$ for different $K^+$ polar angle bins. The bands are 
%projections  by ANL-Osaka (red), BnGa 2014 (green), and BnGa 2016 (blue). The latter included the new data in the fit. 
%The large discrepancies at $W > 2$GeV indicate possible high mass resonance strength that was not included in earlier fits. }
%\label{KLambda_Oz}
%\end{figure}
%\vspace{-2.5cm}

\subsection{Single pion and vector meson photoproduction}
Single pion production used to be the workhorse in the search for new baryon states in the light quark sector, and in particular
in the elastic $\pi N \to \pi N$. The fact that very few high mass states were found in the elastic channel  led to 
the "missing resonance" crises and the move away from this channel and to photoproduction processes. With the high precision possible 
and with the use of polarization as an effective tool, processes $\gamma N \to \pi N$ may still be an effective process even 
for single pion production if the initial state channel $\gamma N \to N^*$/$\Delta$ has significant photocoupling amplitudes.
Figure~\ref{Delta2200} shows the example of the $\Delta(2200){7\over 2}^-$, the parity partner of 
$\Delta(1950){7\over 2}^+$. The latter is situated at much lower mass than the former. The $\Delta(2200){7\over 2}^-$ is a candidate state with 
poor evidence and a low 1* rating the RPP~\cite{Olive:2016xmw}. Precise measurements of  cross sections and single and double 
polarization observables have sufficient sensitivity to uniquely identify the 
state~\cite{Anisovich2017} as shown in Fig.~\ref{Delta2200}. 

Other channels, such as $\gamma p \to p \omega$~\cite{Williams:2009ab,Williams:2009aa} and
 $\gamma p \to \phi p$~\cite{Seraydaryan:2013ija,Dey:2014tfa} are now abundantly 
available, including high statistics polarization measurements. These large and high precision data sets still need to be 
fully utilized in multi-channel partial wave analyses.  

\subsection{Search for hybrid mesons and baryons}

Gluonic excitations of baryons (hybrid baryons) have been explored theoretically and predictions of hybrid baryon 
masses and quantum numbers have recently been made in Lattice QCD. In distinction to the meson sector, where 
some hybrid meson states have "exotic" $J^P$ combinations that are not possible in $q\bar{q}$ quark model 
configuration, hybrid baryons have the same quantum numbers as regular three-quark baryons, and cannot be 
identified through their quantum numbers. The different internal structure of hybrid baryons leads to transition 
amplitudes with a very different $Q^2$ dependence than ordinary $q^3$ baryons. 
For example, for the lowest mass, Roper-like, hybrid baryon with spin-parity $J^P = {1\over 2}^+$, the scalar 
amplitude is predicted to be zero and the transverse amplitude $A_{1/2}$ should drop rapidly with $Q^2$, 
while both is not expected for the equivalent $q^3$ state. These circumstances are exploited in the planned  
experimental program with CLAS12 at JLab,   
that uses even higher luminosities than previously available to measure several reactions, such as 
$K^+\Lambda$, $N\pi$, $p\pi^+\pi^-$, and others, to search for new states in the mass range from 2.0 to 
3.0 GeV and to measure their transition helicity amplitudes at low $Q^2$, where hybrid baryon contributions 
are expected to be more prominent.  The hybrid meson program at 12 GeV has led to the development of a 
quasi-real photon tagger facility for CLAS12~\cite{e12-11-005} that can also be utilized in the search for new baryon states in that high
mass region, and measure their transition form factors~\cite{e12-16-010}. The linear polarization of the virtual photon 
plays an important role in both programs to identify the quantum numbers of excited states, while measurement
of the transition amplitudes help in identifying the hybrid nature of baryons.   
%\vfill\eject

\section{Polarized Parton Distributions} 
\label{pdfs}
\begin{figure}[h]
\vspace{-0.7cm}\includegraphics[width=8.5cm,height=8.5cm]{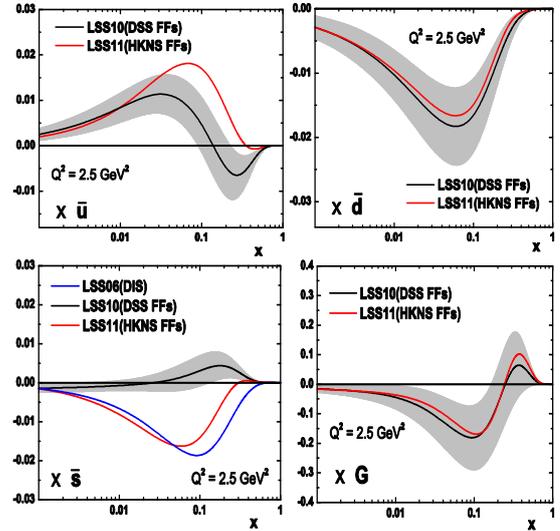}
\caption{Current status of the sea-quark and gluon distribution functions from DIS and SIDIS data 
(see: talk by E. Leader)}
\label{Leader_16}   
\end{figure}

Polarized quark distribution have been studied over more than 2 decades and excellent information is now available on 
the valence quark $u(x)$ and $d(x)$ distribution. On the other hand the sea quark distributions and the gluon distribution
are much less well known. Figure~\ref{Leader_16} shown by Elliott Leader at this 
conference, highlights the results of one of the global analysis of inclusive scattering data. The uncertainties in the data still
leave significant room for improvement. One of the discrepancies in $x\Delta{s}$ that has been resolved is the sign 
change difference of the DIS with the analysis 
using SIDIS data with specific fragmentation function (FF). Using a different set of FF brings the DIS and DIS+SIDIS results
in agreement. Also, the gluon distribution $x\Delta{G}$ has still very large uncertainties. 

\begin{figure}[t]
\hspace{-1.0cm}\includegraphics[width=9.5cm,height=8cm]{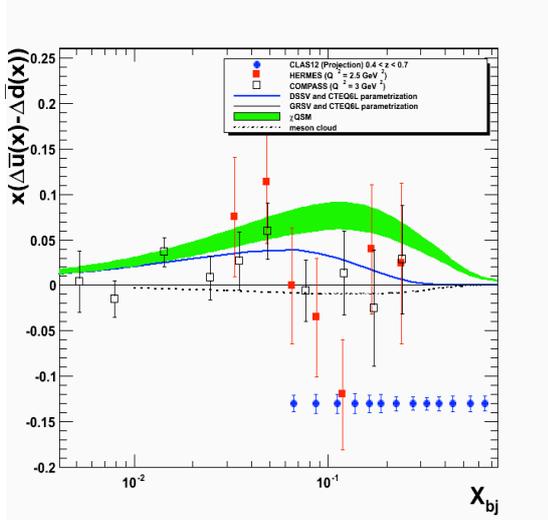}
\caption{ Projected kinematic range and uncertainties (full circles) for the asymmetry of 
the polarized sea using polarized electron beams on both polarized $NH_3$ and 
polarized $ND_3$ targets, using CLAS12.}
\label{pol-sea-asym}    
\end{figure}

Precise information is also lacking on the spin structure function $g_1(x,Q^2)$ 
and the helicity asymmetry $A_1(x)$ at large values of $x$. Several experiments 
~\cite{e12-06-109,e12-06-110,e12-06-122} will study polarized parton distributions 
at large $x$ using polarized $NH_3$, $ND_3$, and $^3He$ as target material. 
Examples of the improvement expected for the polarized d-quark density and the 
asymmetry of the polarized sea are shown in Fig.~\ref{pol-sea-asym}.

Parton distribution functions are extracted from experimental data in 
global analysis that make use of all available data. The limited information 
available for the polarized structure function $g_1(x,Q^2)$ has as a consequence 
that the polarized gluon density $\Delta G(x)$ is the least constrained of the 
parton distribution function. The precision of the expected data on $g_1(x,Q^2)$ 
at 12 GeV provides a model-independent way of determining $\Delta G(x)$ through
the $Q^2$ dependence measured in a large range of $Q^2$. The improvement
from the 12~GeV upgrade is 
also significant at lower $x$. Other polarized parton densities, especially 
$\Delta s$ will be improved as well using polarized proton and deuterium targets. 
The projected data will not only reduce the error band on $\Delta G$, but will likely 
allow a more detailed modeling of its $x$-dependence. Fig.~\ref{pol-sea-asym} shows
the expected uncertainty for the asymmetry of the polarized sea from measurements
with polarized proton ($NH_3$) and deuterium ($ND_3$)targets.

\section{3D Imaging of the proton and orbital motion}
Recent LQCD calculations~\cite{Alexandrou:2016mni,Liu:2015nva} of the spin decomposition in the proton, 
for the first time have included the disconnected interactions (DI). The results have spectacularly
overturned previous LQCD calculations that need not have the DI included. Instead of obtaining 
negligible net orbital angular momentum (OAM) contributions,  net OAM of $\approx 45\%$  have
 been obtained when DI are included. These results make the broad planned GPD and TMD 
 programs at Jefferson Lab all the more important and urgent as the possibly largest contribution 
 to the proton spin has not been experimentally accessed. These calculations start from the 
 gravitational form factors of the nucleon matrix element of the energy-momentum tensor, and 
 compute the elements of the 
 gauge invariant decomposition $$J_p = ({1\over 2}\Delta{\Sigma} + L^q) + G^g, $$
 where $\Delta{\Sigma}$ is the quark helicity, $L^q$ is the quark OAM,  and $G^g$ is the 
 gluon spin contribution. The quark contribution $L^q$ is accessible through measurement of 
 the generalized parton distributions, which discussed in the following section. Details are discussed
 in section~\ref{gpds} and section~\ref{gff}.

\subsection{Generalized Parton Distributions and DVCS}
\label{gpds}
\begin{figure}[t]
\hspace{-0.5cm}
\includegraphics[width=9.0cm,height=8.5cm]{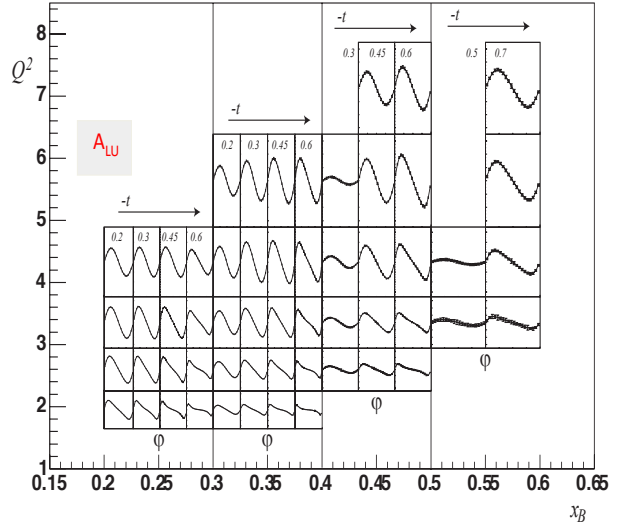}
\caption{Projected data for the beam spin asymmetry $A_{LU}$ 
for the DVCS-BH interference of experiment ~\cite{e12-06-119}. 
Statistical uncertainties are too small to be visible.} 
\label{alu_dvcs}
\end{figure}
\begin{figure}[t]
\vspace{-0.5cm}\hspace{-0.8cm}\includegraphics[width=8.cm,height=8.cm]{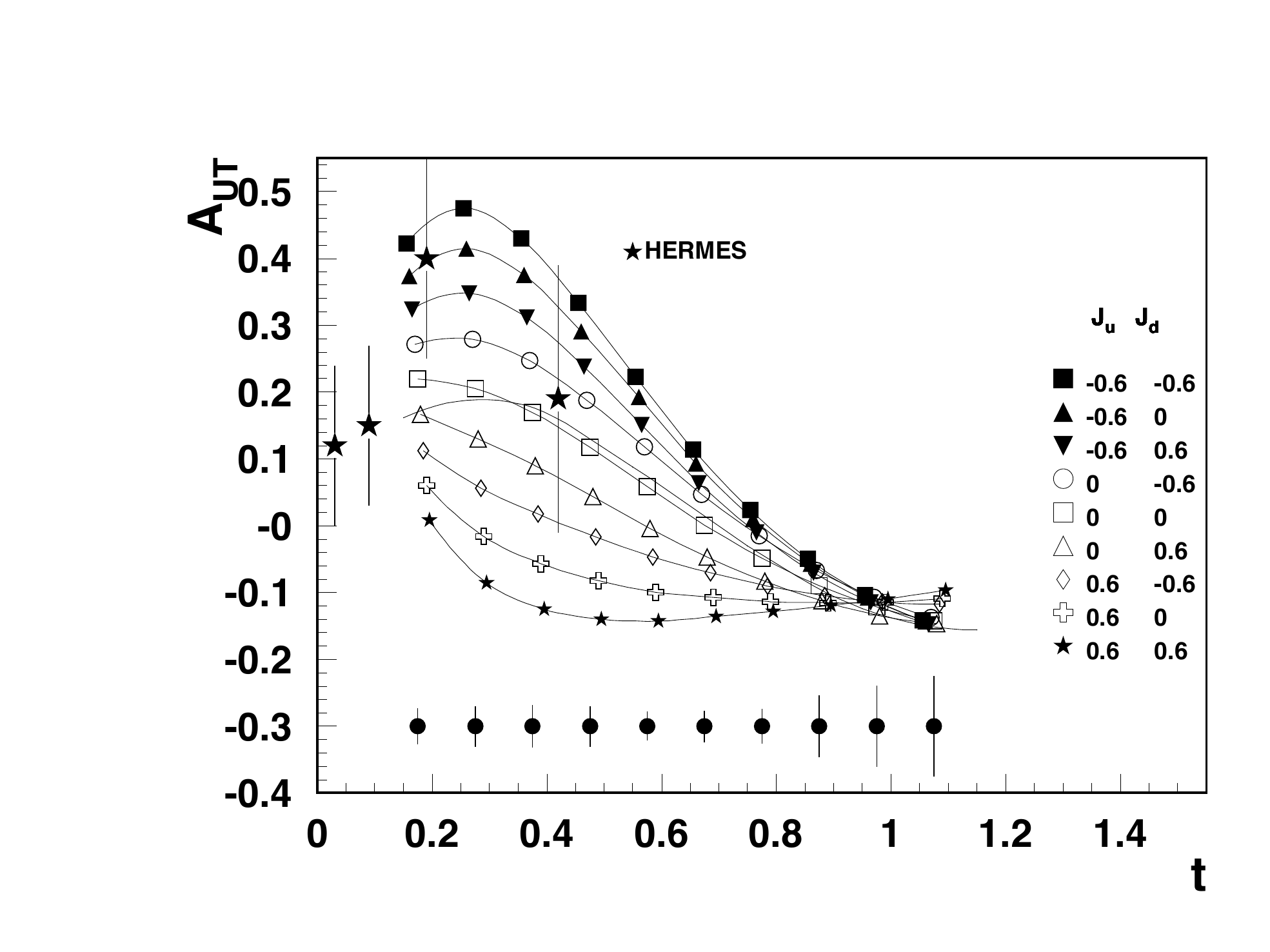}
\caption{Projected data for the transverse target asymmetry for the 
proposed experiment ~\cite{e12-12-010}. The $t$ dependence 
of $A_{UT}$ is shown for a single bin in $Q^2$ and $x$. The  projections are 
based on the use of a polarized HD target in CLAS12.}
\label{aut_dvcs}    
%\end{figure}
%\begin{figure}[h]
\hspace{-2.0cm}\includegraphics[width=11cm,height=8cm]{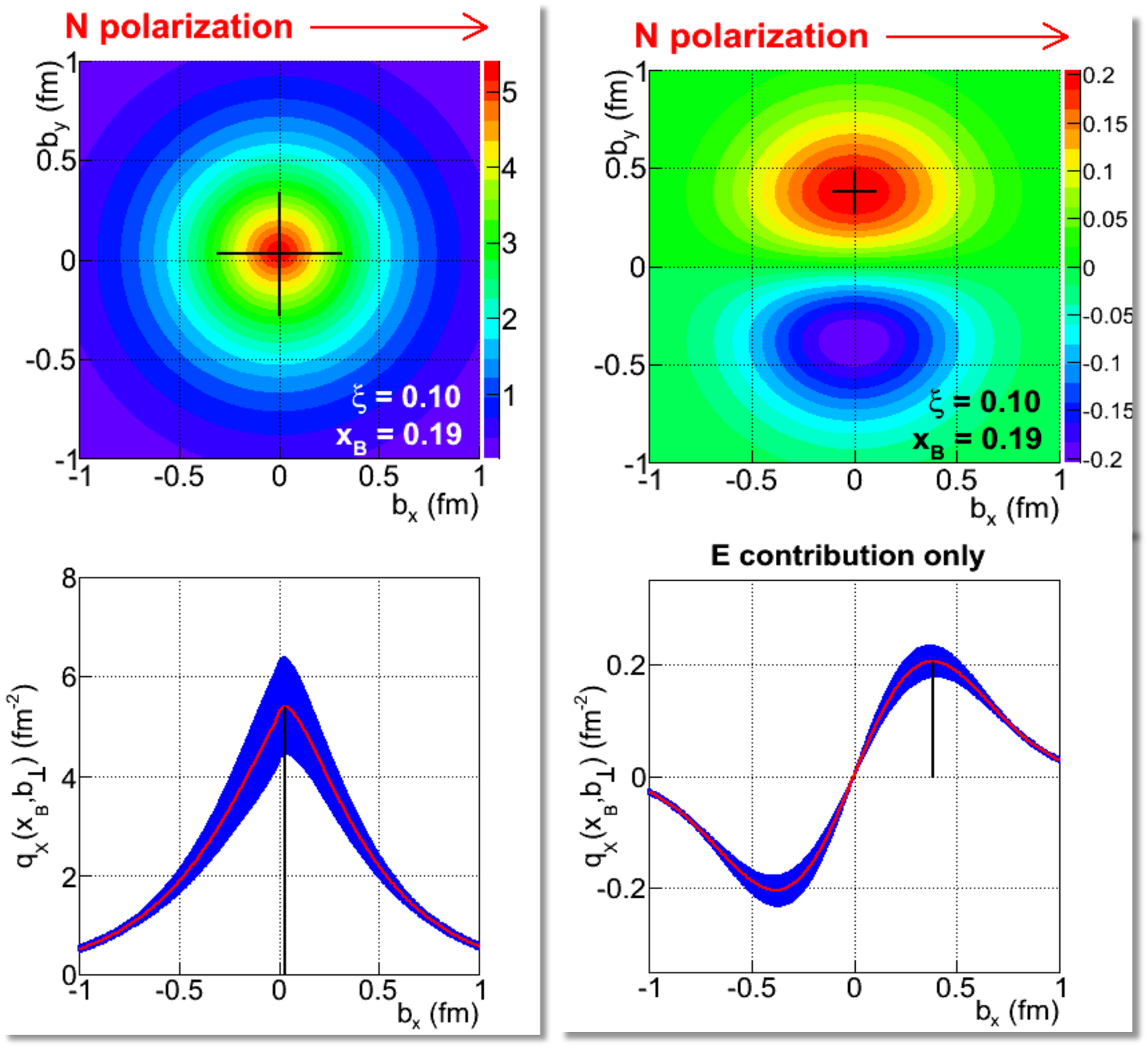}
\caption{Top: Projected quark distribution in transverse impact space as determined from 
GPD $H + E$ (left) and from GPD $E$ alone (right). Bottom: 
Projections on $b_y$ axis. Distribution of $E$ clearly shows the transverse shift expected from
internal orbital angular momentum (graphics courtesy of F.X. Girod). The vertical shift of the 
center in the left graph is a measure of the size of the OAM contribution to the proton spin.   }
\label{GPD_H_E}    
\end{figure}
\noindent At the lower and intermediate energies DVCS is the most direct way to access GPDs, and
 information on GPDs has recently been extracted from JLab 
and Hermes data~\cite{Guidal:2013rya}. DVCS is most suitable for
mapping out the twist-2 vector GPDs $H, E$ and the axial GPD 
${\tilde H}$. Having access to a 3-dimensional image
of the nucleon opens up new insights into the complex structure of the
nucleon. The beam helicity-dependent cross section asymmetry is given 
in leading twist as 
$$ A_{LU} \approx \sin\phi[F_1(t)H + \xi(F_1+F_2)\tilde{H}]d\phi~, $$where
$\phi$ is the azimuthal angle between the electron scattering plane and the hadronic plane. 
The kinematically suppressed term with GPD $E$ is omitted. 
The asymmetry is mostly sensitive to the GPD $H(x=\xi,\xi,t)$. 

The target asymmetry 
$A_{UL}=\Delta\sigma_{UL}/2\sigma$ accesses the same GPDs but emphasizes GPD $\tilde H$ 
more strongly, while GPD $H$ is kinematically unfavored, although at not 
too small values of $\xi$, $H$ and $\tilde H$ both contribute, and the combination 
of $A_{LU}$ and $A_{UL}$ enables the separation of GPD $H(x=\xi,\xi,t)$ and
$\tilde{H}(x=\xi,\xi,t)$.  
Using a transversely polarized target the asymmetry 
$$A_{UT} \approx \cos\phi\sin(\phi-\phi_s) [t/4M^2 (F_2H - F_1 E)] $$ can be measured, 
where $\phi_s$ is the azimuthal angle of the target polarization vector relative to the electron scattering plane. $A_{UT}$ 
depends in leading order on GPD $E$. 

The first DVCS experiments carried out at JLab~\cite{Stepanyan:2001sm,Girod:2007aa,Camacho:2006qlk,Chen:2006na} 
and DESY~\cite{Airapetian:2001yk} showed promising 
results in terms of the applicability of the handbag mechanism to probe GPDs. The 12 GeV upgrade offers 
much improved possibilities for accessing GPDs. 
Figure~\ref{alu_dvcs} shows the expected statistical precision of 
the beam DVCS asymmetry for the full kinematics. Using polarized targets one can measure 
the  target spin asymmetries with high precision. Figure~\ref{aut_dvcs} shows 
the expected statistical accuracy for one kinematics bin of the transverse target asymmetry. 
A measurement of all 3 asymmetries will enable the separation of GPDs 
$H,~\tilde{H}$ and $E$ at the above specified kinematics. A Fourier transformation
of the t-dependence of GPD $H$ can be used to determine the quark distribution 
in transverse impact parameter space. From the $t$ dependence of GPD $E(x,t)$ 
one obtains the angular momentum distribution in transverse impact parameter. 
Figure~\ref{GPD_H_E} shows projected results for the transverse impact parameter
distribution of quarks in transversely polarized target using a model for GPDs and 
the expected DVCS data at 11 GeV beam energy with all unpolarized cross sections and 
all polarization configurations measured.  
\vfill
\subsection{Gravitational form factors and confinement.} 
\label{gff}
In addition to providing the basis for spatial imaging of the proton GPDs carry information of 
more global nature. For example, the nucleon matrix element of the energy-momentum tensor 
contains 3 form factors that encode information on the angular 
momentum distribution $J^q(t)$ of the quarks with flavor $q$ in transverse space, their 
mass-energy distribution $M_2^q(t)$, and their pressure and force 
distribution $d^q_1(t)$. These form factors appear as moments of the vector 
GPDs~\cite{Goeke:2007fp}, thus offering
prospects for accessing gravitational form factors through the detailed mapping of GPDs. 
 As an example, 
the quark angular momentum distribution in the nucleon is given by 
$$J^q(t) - 4/5d^q_1(t)\xi^2 = \int_{-1}^{+1}dx x E^q(x, \xi, t)~,$$
and the mass-energy and pressure distribution $$M_2^q(t) + 4/5d^q_1(t)\xi^2 
= \int_{-1}^{+1}dx x H^q(x, \xi, t)~.$$ The mass-energy and force-pressure distribution 
of the quarks are given by the second moment of GPD $\it{H}$, and their relative 
contribution is controlled by $\xi$. 

For the extraction of GPD $H$, the electron beam must have spin aligned parallel and anti-parallel to the beam 
direction to determine $A_{LU}$, and the analysis uses dispersion relations to determine 
$d^q_1(t)$. A Fourier transform in $t$ allows for the determination of the radial pressure distribution 
using a parameterization for the $t$ dependence of the form factor. An experimental program has been 
approved~\cite{e12-16-010b} to obtain the high statistics data at several beam energies that are needed 
for this analysis.     

\begin{figure}[t]
\vspace{-1cm}\hspace{-0.5cm}\includegraphics[width=8.5cm,height=8.5cm]{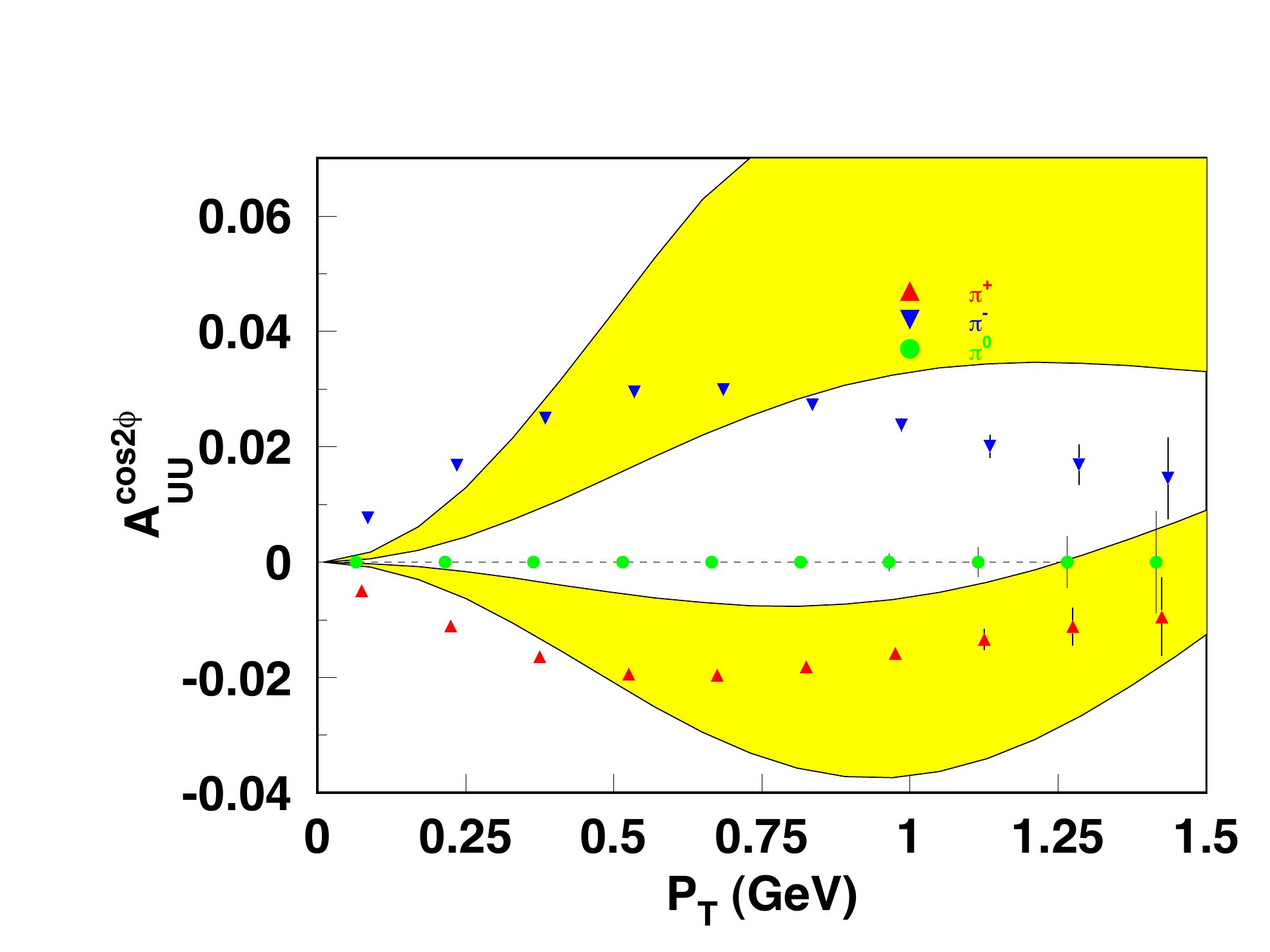}
\caption{The $\cos2\phi$ moment (Boer-Mulders asymmetry) for pions
as a function of $Q^2$ and $P_T$ for $Q^2>2$~GeV$^2$ (right) with {\tt CLAS12} 
at 11~GeV from 2000~hours of running.  Values are calculated assuming
$H_1^{\perp u\rightarrow \pi^+}=-H_1^{\perp u\rightarrow \pi^-}$. Only statistical
uncertainties are shown.}
\label{fig:Boer-Mulders}    
\end{figure}   

\subsection{TMDs and semi-inclusive DIS}
\label{tmds}
\noindent Semi-inclusive deep inelastic scattering (SIDIS) processes give access to the 
transverse momentum distributions of quarks (TMDs). Azimuthal distributions of final state particles in SIDIS  
provide access to the orbital motion of quarks and 
play an important role in the extraction of TMDs of quarks in the nucleon. 
\begin{table}[h]
\caption{{{\rm Leading-twist transverse momentum-dependent distribution 
functions.  $U$, $L$, and $T$ stand for transitions of unpolarized, 
longitudinally polarized, and transversely polarized nucleons (rows) to 
corresponding quarks (columns).}}\label{tab1}} 
\hspace{2cm}
\begin{tabular}{|c|c|c|c|} \hline\hline
N/q & U & L & T \\ \hline
 {U} & ${\bf f_1}$   & & ${ h_{1}^\perp}$ \\ \hline
 {L} & &${\bf g_1}$ &    ${ h_{1L}^\perp}$ \\ \hline
 {T} & ${ f_{1T}^\perp} $ &  ${ g_{1T}}$ &  ${ \bf h_1}$ \, ${ h_{1T}^\perp }$ \\
\hline\hline
\end{tabular}
\end{table}
TMD distributions describe transitions of a nucleon 
with one polarization in the initial state to a quark with another polarization 
in the final state. The diagonal elements of the table~\ref{tab1} are the momentum, 
longitudinal and transverse spin distributions of partons, and represent well-known parton
distribution functions related to the square of the leading-twist, light-cone 
wave functions. Off-diagonal elements require non-zero orbital angular 
momentum and are related to the wave function overlap of $L$=0 and $L$=1 Fock 
states of the nucleon.  The chiral-even distributions 
$f_{1T}^\perp$ and $g_{1T}$ are the imaginary parts of the corresponding
interference terms, and the chiral-odd $h_1^\perp$ and $h_{1L}$ are the
real parts.  The TMDs $f_{1T}^\perp$ and  $h_{1}^\perp$, which are related to 
the imaginary part of the interference of wave functions for different orbital 
momentum states and are known as the Sivers and 
Boer-Mulders functions. They describe unpolarized quarks in the 
transversely polarized nucleon and transversely polarized quarks in the 
unpolarized nucleon respectively.  

The most simple mechanism that can lead to a Boer-Mulders function is a 
correlation between the spin of the 
quarks and their orbital angular momentum.  In combination with a final state 
interaction that is on average attractive, already a measurement of the sign 
of the Boer-Mulders function, would thus reveal the correlation between 
orbital angular momentum and spin of the quarks.

\begin{figure}[h]
 \includegraphics[width=8cm,height=8cm]{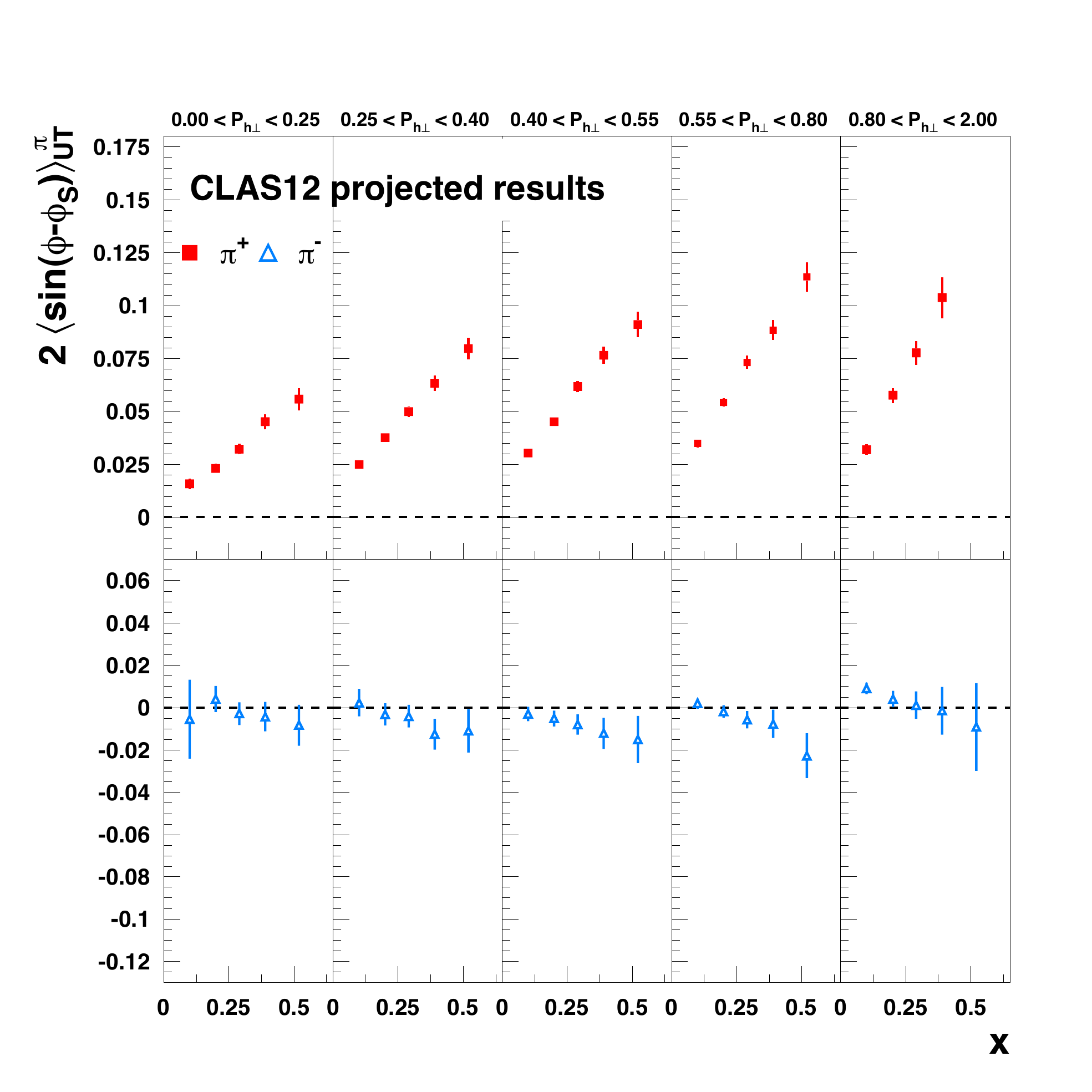}
\caption{Projected data for the Sivers function vs $x$ for different transverse momentum bins (left panel) and vs $p_T$ for different $x$ bins. }
\label{sivers}    
\end{figure}

Similar to the GPDs, TMD studies will greatly benefit from the higher energy and high 
luminosity at 12 GeV. A comprehensive program is in preparation at JLab to study  
new structure functions that encode the TMDs. The main focus is the measurement of SIDIS processes in 
the full phase space available. Examples of expected uncertainties~\cite{e12-06-112} for 
the Boer-Mulders asymmetry $A^{cos2\phi}_{UU}$ in SIDIS of $\pi^+$, $\pi^0$, and $\pi^-$  
are presented in Fig.~\ref{fig:Boer-Mulders}. Projected data for the $x$ dependence of 
the Sivers asymmetries for $\pi^+$ and $\pi^-$ are shown in Fig.~\ref{sivers} for bins in 
the pion transverse momentum.  

While pions are currently the main focus of the JLab TMD program, SIDIS with K$^+$ or
 K$^-$ as leading particles are also of high interest. Earlier HERMES results show unexpectedly 
 large Boer-Mulders asymmetries for kaons compared to pions, and the opposite signs 
 for K$^-$ and $\pi^-$.   With the 
 excellent particle identification and high luminosity expected for CLAS12, these puzzling
 issues can be addressed very efficiently.  

\section{Conclusions}
\label{summary}
The JLab 12 GeV energy upgrade and the new experimental equipment 
are well matched to an exciting scientific program aimed at studies of the complex nucleon 
structure in terms of the  longitudinal and transverse momentum dependent parton distribution 
functions, the GPDs and TMDs. They provide fundamentally new insights into the complex and 
multi-dimensional internal structure of the nucleon. For the first time, direct information on 
the orbital angular momentum of quarks in protons may become accessible experimentally, as
does  the distribution of confinement forces acting on the quarks in the proton. Other programs 
will extend the spin structure function measurements of protons and neutrons to higher $x$ and 
search for new meson and baryon states with significant components of "glue" in their wave functions.  
The CLAS12 detector will be at the core of this exciting program.    

\section*{Acknowledgement}
I like to thank F.X. Girod for the revealing graphics in Fig.~\ref{GPD_H_E}. 
The material presented here is based upon work supported by the U.S. Department of Energy, Office of Science, 
Office of Nuclear Physics  under  contract DE-AC05-06OR23177.

\end{document}